\documentclass[aps,prl,twocolumn,showpacs,superscriptaddress,groupedaddress,floats,rmp,amsmath]{revtex4-1}

\usepackage{graphics,graphicx,epsfig}
\usepackage{amssymb}
\usepackage{epstopdf}
\usepackage{color}
\usepackage[usenames,dvipsnames]{xcolor}

\begin{document}


 \title{Maternal origins of developmental reproducibility}

\author{Mariela D. Petkova$^{a}$}
\author{Feng Liu$^{a}$}
\author{Thomas Gregor$^{a,b}$}

\affiliation{$^a$Joseph Henry Laboratories of Physics, and
$^b$Lewis-Sigler Institute for Integrative Genomics, \hfill
Princeton University, Princeton, New Jersey 08544 USA}

\date{\today}

\begin{abstract} 
Cell fate decisions in multicellular organisms are precisely coordinated, leading to highly reproducible macroscopic outcomes of developmental processes. The origins of this reproducibility can be found at the molecular level during the earliest stages of development when spatial patterns of morphogen (form-generating) molecules emerge reproducibly. However, the initial conditions for these early stages are determined by the female during oogenesis, and it is unknown whether reproducibility is passed on to the zygote or whether it is reacquired by the zygote. Here we examine the earliest reproducible pattern in the Drosophila embryo, the Bicoid (Bcd) protein gradient.  Using a unique combination of absolute molecule counting techniques, we show that it is generated from a highly controlled source of mRNA molecules that is reproducible from embryo to embryo to within $\sim\!8\%$. This occurs in a perfectly linear feed-forward process:  changes in the female's gene dosage lead to proportional changes in the mRNA and protein counts in the embryo. In this setup, noise is kept low in the transition from one molecular species to another, allowing the female to precisely deposit the same absolute number of mRNA molecules in each embryo and therefore confer reproducibility to the Bcd pattern. Our results indicate that the reproducibility of the morphological structures that emerge in the embryo originates during oogenesis when all initial patterning signals are controlled with precision similar to what we observe for the Bcd pattern. \end{abstract}


\maketitle

\section{Introduction}

Macroscopic structural patterns of multicellular organisms are reproducible from one individual to the next~\cite{Thompson:1917, Maynard-Smith:1960}. Insects, in particular, show strikingly reproducible morphological features: patterns of body segments and locations of hair bristles  are essentially identical in individuals within a species ~\cite{Maynard-Smith:1960, Wigglesworth:1940, Lawrence:1973}. The spatial locations of these features are controlled by underlying molecular regulatory networks, which face stochastic noise \cite{Arias:2006, Lander:2007, Wolpert:2007}. It is unclear how exactly these networks cope with molecular fluctuations to achieve reproducible outcomes, but two distinct strategies have been conceptualized {~\cite{Schrodinger:1944, Bialek:2012}}. In the first case, the initial and boundary conditions are set up reproducibly and all subsequent processes have been tuned to minimize noise. Alternatively, patterns start out noisily and error-prone, but special error-correcting mechanisms minimize the effects of noise as development proceeds (Fig.~1). These two opposing cases lead to fundamentally different perspectives on how to experimentally and theoretically examine developmental signaling networks and it is key to identify systems in which such differences can be exposed.  

\begin{figure}[t]
\centering
\includegraphics[width=\linewidth]{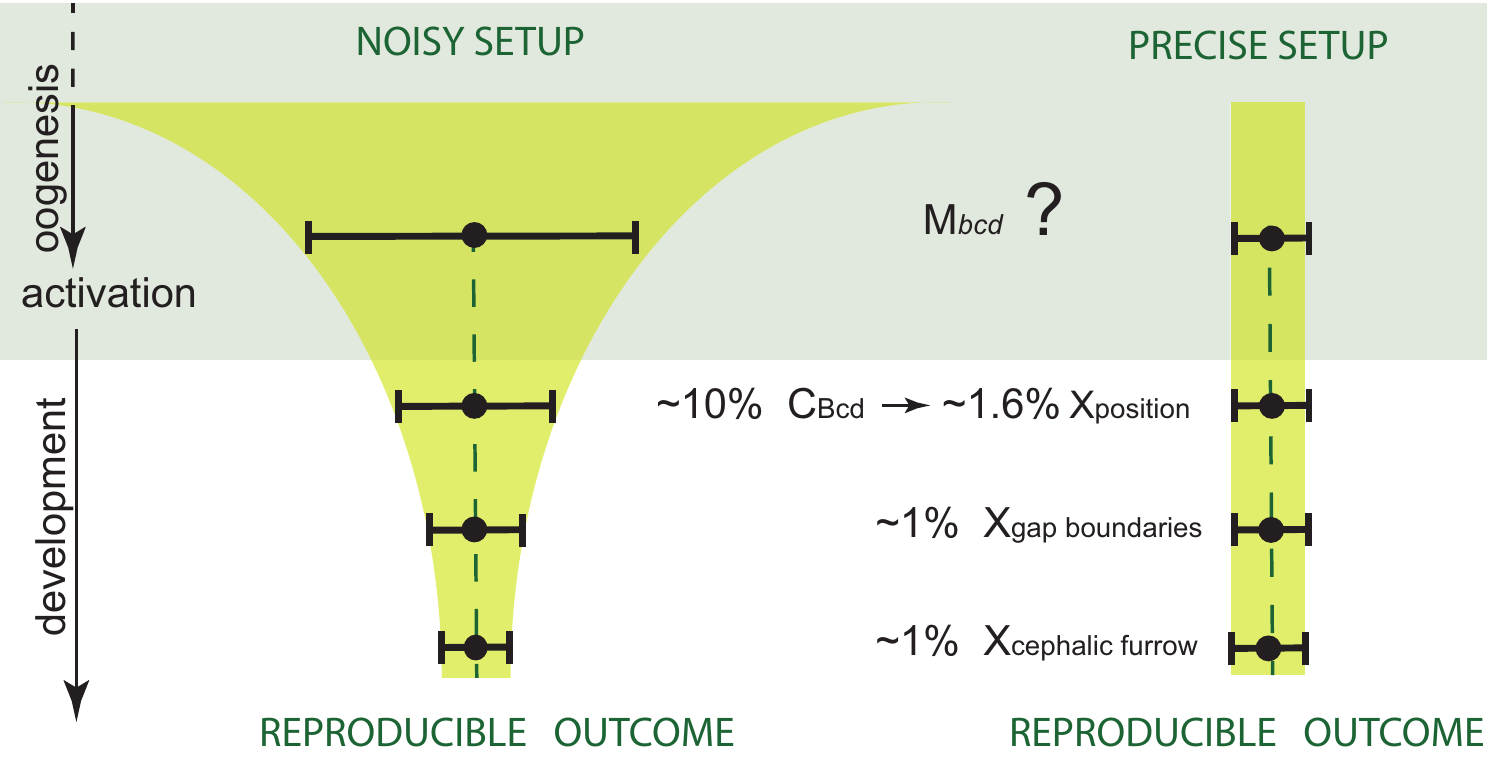}
\caption{Two possible scenarios with reproducible developmental outcomes. In the ``funnel" scenario, large fluctuations in the initial setup are reduced as development proceeds. In the ``channel" scenario, at each step along the developmental cascade reproducible outputs originate from reproducible inputs. In {Drosophila} embryos, the Bicoid (Bcd) protein concentration gradient $C_{\rm Bcd}$ is reproducible with $\sim\!\!10\%$ accuracy. This reproducibility translates into a spatial accuracy of $1\!-\!2\%$, i.e. the size of a single cell, sufficient to induce distinct fate decisions. In fact, locations of pattern features (denoted by $X$) along the developmental cascade demonstrate similar reproducibility. We show that the fly mother confers the reproducibility of the initial Bcd pattern to the embryo by assembling a reproducible mRNA source $M_{bcd}$ during oogenesis.} 
\label{fig0}
\end{figure}

Embryos of the fruit fly {\it Drosophila melanogaster} provide a unique opportunity to quantitatively study how molecular processes are orchestrated to achieve reproducible outcomes. Cells along the anterior-posterior (AP) axis of the developing embryo determine their location by interpreting spatial patterns of morphogen molecules whose concentration correlates with AP position. The processes which lead to the expression of these molecular patterns (reviewed in Ref. \cite{Jaeger:2011}) begin in the female during oogenesis when particular mRNA molecules are localized at the cardinal points of the egg. Upon egg activation, these molecules serve as sources for protein gradients which trigger a network of interacting genes that generate a cascade of increasingly diversified molecular spatial patterns that eventually specify unique fates for each of the $\sim\!80$ rows of cells along the AP axis~\cite{Gergen:1986}. Specifically, one of the first patterns is the Bicoid (Bcd) protein gradient whose diffusion-driven exponentially decaying concentration profile along the AP axis~\cite{Crick:1970, Driever:1988a} is reproducible to within $10\%$ from embryo to embryo~\cite{Gregor:2007a,Gregor:2007b}. This reproducibility is sufficient for the Bcd gradient to encode position with 1.6\% embryo length (EL) precision. In fact such spatial precision is observed in positioning of features of the subsequent molecular patterns~\cite{ Crauk:2005, Surkova:2008,Dubuis:2013,Liu:2013} that lead up to the first macroscopic structure in the embryo, the cephalic furrow, which emerges  when a single row  of cells located at $(33\pm1)\%$EL folds inward~\cite{Namba:1997, Vincent:1997}.

Hence, the early patterning cascade in the developing embryo seems to follow a maximally reproducible path: from the setup of the input gradient to its establishment and readout at consecutive levels, all the way to first macroscopic morphological features such as the cephalic furrow (Fig.~1). However, it is unknown whether reproducibility is passed onto the embryo by the female or whether the embryo must acquire reproducibility via specialized error-correcting mechanisms. Here we address this question by examining the assembly of the mRNA source for the Bcd protein gradient. A priori, it is possible that the number of source mRNA molecules fluctuates significantly from embryo to embryo. In this case one expects the presence of error-correcting mechanisms, such as feedback regulation, which generate the observed reproducibility at the protein level~\cite{Eldar:2003, England:2005, Wojcinski:2011}. Alternatively, the female deposits {\em bcd} mRNA with $\sim$10\% precision and all embryos control the kinetics of protein synthesis identically. In this case, no special error-correcting mechanisms are necessary. Instead, the reproducibility of the initial pattern is set by the precision with which the female assembles the source mRNA molecules from embryo to embryo. Intriguingly, our measurements are in tune with the latter possibility.

%
%
%
%

\section{Results}

\noindent {\bf Maternally Deposited mRNA Are Reproducible at the Same Level as Zygotic Patterns.} 
Tracing the origin of developmental reproducibility in the fly embryo leads to the question whether the reproducibility of the initial patterns is reflected already in their setup, i.e. the maternally deposited mRNA cues. Is the female able to count mRNA molecules during oogenesis? To answer this question we need to be able to count absolute numbers of mRNA molecules in individual embryos. Here we devised an experimental strategy that counts {\em bcd} mRNAs by two independent means.

\begin{figure}[t]
\centering
\includegraphics[width=\linewidth]{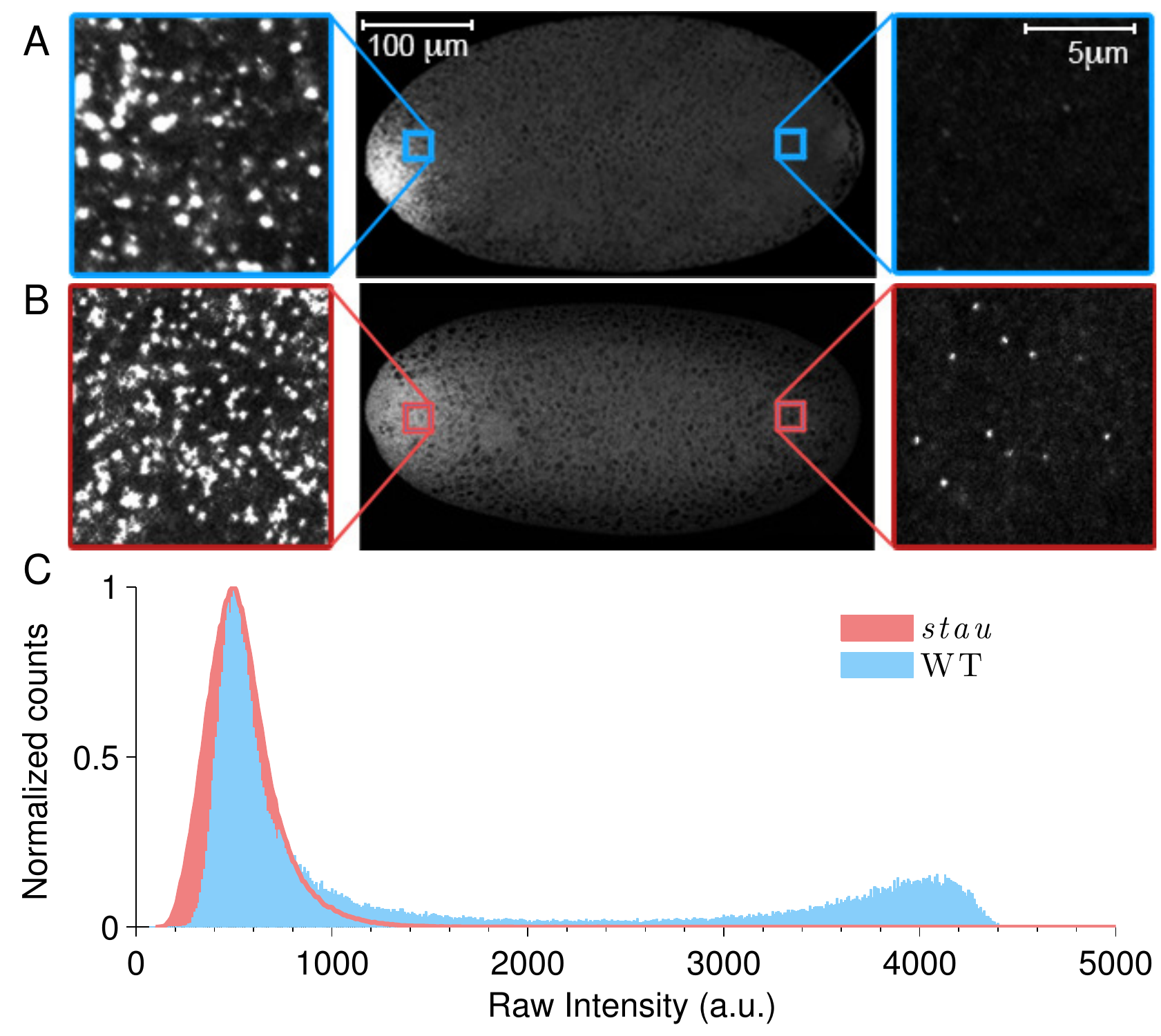}
\caption{{\it In situ} total mRNA measurements using single molecule counting. \emph{bcd} mRNA molecules in fixed embryos are tagged with 90 fluorescently labeled oligonucleotide probes (20-mers) for wild-type {\bf(A)} and \emph{stau$^-$} mutant {\bf(B)} embryos. Packages of multiple \emph{bcd} mRNA molecules dissociate in \emph{stau$^-$} mutants (side panels), as apparent from unimodal particle intensity distribution. {\bf(C)} Distributions of raw fluorescence  (in arbitrary intensity units) per particle for wild-type and \emph{stau$^-$} embryos. Particles in the low intensity peaks are identified as single molecules using a threshold-based detection (SI Materials and Methods, Fig~S1). Fluorescence values across experiments differ by a constant background fluorescence, and for comparison, the distributions have been aligned at their single molecule peaks. }
\label{fig2}
\end{figure}

First, we extended a recently developed mRNA labeling and counting method, based on fluorescence \emph{in situ} hybridization (FISH)~\cite{Little:2011}, to optically identify individual mRNA molecules in whole mount embryos and assess their embryo-to-embryo reproducibility~\cite{Little:2013}. However, wild-type embryos labeled with this technique reveal  packaged {\em bcd} mRNA bundles (Fig.~2A) that are held together by a protein called Staufen (Stau)~\cite{Ferrandon:1994}. In embryos that lack the {\it stau} gene ({\it stau$^-$}) these packages are resolved into individual mRNA particles, which we can detect reliably using confocal microscopy (Fig.~2B). The particle intensity distribution in {\it stau$^-$} embryos is unimodal as opposed to bimodal in wild-type embryos (Fig.~2C), suggesting that particle packages have indeed been resolved. We measure a total count of $M_{bcd}^{\rm FISH}=(9.0 \pm0.7)\times10^5$  {\em bcd} mRNA molecules in individual {\it stau$^-$} embryos, where the error is the standard deviation across $n\!=\!7$ embryos (SI Materials and Methods). Strikingly, the embryo-to-embryo reproducibility is ${8\pm2\%}$ (bootstrapping error) in {\em bcd} mRNA counts which is of the same order as that of Bcd proteins (${\sim10\%}$ ~\cite{Gregor:2007b}). Hence, we conclude that the female has an exquisite control over the mRNA source composition, deploying mRNA with sufficient precision to generate a reproducible Bcd gradient without the need of a special error-correcting mechanism.

This FISH-based approach harbors two major potential pitfalls. First, the conversion from optically detected mRNA particles to mRNA molecules $M_{\it bcd}$ is non-trivial; it is possible that each of the individual mRNA particles we detect in {\it stau$^-$} embryos is in fact comprised of multiple mRNA molecules. Second, {\it stau}$^-$ females might produce different mRNA numbers than wild-type females, in which case the conclusions made above cannot be extended to wild-type flies. To directly address these issues, we developed an mRNA counting method based on bulk mRNA measurements in whole embryos.

\vspace{.5cm}
\noindent {\bf Double-Calibrated Absolute qRT-PCR Confirms Total mRNA Counts.} 
To confirm that we optically detect individual mRNA molecules and that {\it stau}$^-$ females deploy similar number of mRNA as wild-type females, we developed a second, independent counting approach based on a widely used polymerase chain reaction (PCR) technique~\cite{Bustin:2004}. Quantitative real-time-PCR (qRT-PCR) measures the absolute, total number of mRNA molecules in bulk samples, i.e. homogenized embryos. mRNA molecules are chemically extracted from the sample, converted to DNA by reverse transcription, and  subsequently quantified by fluorescent amplification. 

\begin{figure}[t]
\centering
\includegraphics[width=\linewidth]{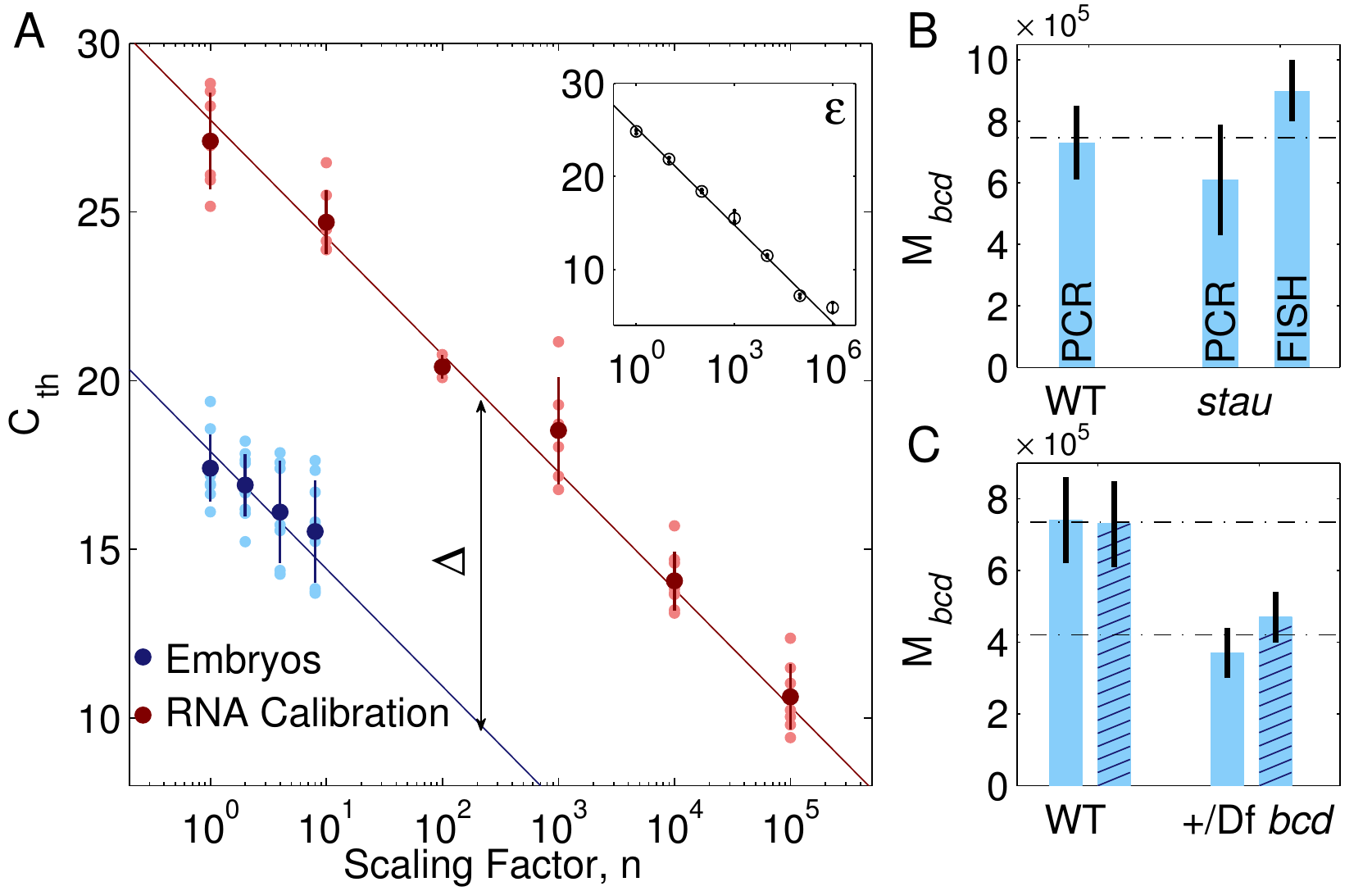}
\caption{Bulk total mRNA counts in single embryos using qPCR. ({\bf A}) mRNA is extracted in bulk from individual samples (colored data points), converted to cDNA and fluorescently amplified via PCR to allow for quantification. Threshold cycle $C_{\rm th}$refers to the PCR amplification cycle for which the sample fluorescence reaches an arbitrary predefined threshold at which all samples are compared. DNA standard curve (inset) is constructed using DNA dilution series to measure PCR amplification efficiency $\varepsilon=10^{-1/{\rm slope}}$ (error bars are smaller than data markers). RNA calibration curve (red) is constructed from samples with $n\times M_{\rm ref}$ synthetically generated mRNA molecules. Embryo target curve (blue) is constructed from samples with n=\{1,2,4,8\} embryos (10\--30min old). The slope of blue and red curves is set by $\varepsilon$, and the number $M_{bcd}$ of total mRNA molecules per embryo is calculated from the offset $\Delta$ between the curves (Materials and Methods). {\bf B)} qRT-PCR measurements of total bcd mRNA in wild-type (WT) and {\it stau$^-$} embryos yield $(7.3\pm1.2)\times10^5$ and $(6.1\pm1.8)\times10^5$ molecules, respectively. With FISH in {\it stau$^-$} embryos $(9.0\pm0.7)\times10^5$ molecules are detected. {\bf (C)} qRT-PCR measurements of total \emph{bcd} mRNA counts in wild-type embryos (2 \emph{bcd} DNA copies) $(7.4\pm1.2)\times10^5$, $(7.3\pm1.5)\times10^5$, and of +/Df \emph{bcd} embryos (1 \emph{bcd} DNA copy) $(4.7\pm0.7)\times10^5$, $(3.7\pm0.7)\times10^5$. Empty and hashed bars indicate independent large-scale experiments and the number of samples in each experiment is found in (SI Materials and Methods, Table~S1)}
\label{fig3}
\end{figure}

Absolute quantification with qRT-PCR is a significant challenge because qPCR results depend highly on the choices of calibrations and reaction conditions~\cite{Bustin:2009}. To date,  conventional qRT-PCR is considered inadequate for measuring absolute mRNA in biological samples, mainly due to inability to quantify the process of RNA isolation~\cite{Nolan:2006}. Conventional methods of absolute mRNA quantification via qRT-PCR use an external calibration curve constructed with synthetic RNA that provides reference to the absolute number of mRNA molecules prior to reverse transcription. This calibration does not control for losses during the extraction process~\cite{Ransick:2004}. To gain such control, we construct a calibration curve from a dilution series of synthetically generated {\em bcd} mRNA molecules that undergo the same procedure in parallel to the homogenized embryo aliquots. We find an extraction efficiency of  $0.3\pm0.1$ (SI Materials and Methods, Fig~S2), demonstrating that the conventional approach underestimates the number of mRNA molecules in embryos at least three-fold.

This ability to correctly calibrate absolute numbers of {\it bcd} mRNA molecules allows us to measure the number of {\em bcd} mRNAs is embryos. Specifically,   we perform qRT-PCR on samples with ${n=\{1,2,4,8\}}$ embryos to construct a target curve that we compare to RNA calibration curve (Fig.~3A).  We also use a dilution series of {\em bcd} DNA molecules to precisely measure the PCR amplification efficiency $\varepsilon$ that determines the slope ($S=-1/\log(\varepsilon)$) of both curves (${\rm CV}(S)<1\%$).   This allows us to perform one-parameter fits with fixed slope $S$ on the target and RNA calibration curves and precisely measure the offset ($\Delta$) between the two curves. The number of mRNA per embryo is then given by $M_{bcd}^{\rm PCR} = M_{\rm ref}\varepsilon^{-\Delta}$. Importantly,  by measuring the amplification efficiency $\varepsilon$ and the offset $\Delta$ with independent calibration curves,  we minimize our experimental error and can calculate it directly using standard error propagation (Materials and Methods).

Using this technique, we obtain $M_{bcd}^{\rm PCR}=(7.3\pm1.2)\times10^5$ {\em bcd} mRNA molecules in embryos from wild-type females,  where the error represents the  measurement error on the mean number of mRNA per embryo (SEM). Note that we measure the mean number of mRNA molecules per embryo with $\sim\!20\%$ precision as estimated from fitting.  In addition, we verify that {\it stau$^-$} embryos contain the same number of {\em bcd} mRNA molecules as wild-type embryos (Fig. 3B). Combining the FISH-based and double-calibrated qRT-PCR molecule counting techniques is key for the internal consistency of our argument. The optical resolution of FISH allows us to count single molecules and measure to embryo-to-embryo mRNA count reproducibility which the large systematic error of the PCR bulk measurements does not allow. On the other hand, because of mRNA packaging in wild-type embryos, single molecule counting is only permissive in {\it stau$^-$} mutant embryos. Here we confirm that the overall number of mRNA molecules in wild-type and mutant embryos are identical, and report an absolute count of {\em bcd} mRNA molecules in individual embryos as the average of the three measurements,  $M_{bcd}=(7.5\pm0.9)\times10^5$ (SEM).

\vspace{.5cm}
\noindent {\bf mRNA and Protein Counts Scale with Maternal Gene Dosage.} 
The apparent exquisite control of the female over the number of mRNA molecules raises the question whether feedback mechanisms act to help assemble the mRNA source reproducibly. On the other hand, the reproducibility of the mRNA source does not rule out the possibility that feedback mechanisms control the protein gradient. To check if feedback mechanisms are involved in controlling either transition from one molecular species to the next, we examine the link between maternal {\em bcd} gene dosage, {\it bcd} mRNA and Bcd protein numbers in the embryo. 

Using our double-calibrated qRT-PCR we can assess how the number of deposited {\em bcd} mRNA molecules scales with maternal {\em bcd} gene copies.  Being diploid, flies carry two alleles (i.e. copies) of each gene in their genome. Using a mutant fly strain in which one allele of the {\em bcd} gene is deleted, we verify that the absolute number of {\em bcd} mRNA molecules in embryos from these {\em bcd} deficient ({+/Df~{\em bcd}}) flies is $0.57\pm0.14$ times that of wild-type embryos with two {\em bcd} alleles (Fig. 3C). The measurement is well within error of the expected factor of 0.5; note that most of the error stems from our measurement capability. Furthermore, it is consistent with previous results obtained using optical approaches that detect a relative difference in mRNA levels ~\cite{Cheung:2011}.

At the protein level we can make a more precise measurement by directly counting Bcd protein molecules in embryos of a set of transgenic fly lines in which the {\em bcd} gene dosage spans a fivefold range~\cite{Liu:2013}. Small changes in maternal {\it bcd} gene expression are achieved by incorporating transgene constructs expressing Bcd-GFP fusion proteins~\cite{Gregor:2007a} at random locations in the fly genome. Insertions at separate locations are expressed at distinct rates which leads to quantitatively different absolute Bcd-GFP distributions and correspond to different apparent {\it bcd} gene dosages (Fig.~4, top inset).  The total number of Bcd-GFP molecules in individual embryos is measured directly by optically calibrating their Bcd-GFP intensity to that of a purified GFP solution of known molarity. Importantly, the embryos are fixed  prior to such measurement to allow all Bcd-GFP proteins to mature and become optically detectable. Applying this method to a reference fly line which expresses Bcd-GFP at wild-type level~\cite{Gregor:2007a,Liu:2013}, we find a total of $N_{\rm Bcd}=(6.7\pm1.5)\times10^7$ (SD) Bcd-GFP molecules in approximately two hour old embryos (SI Materials and Methods).

\begin{figure}[t]
\centering
\includegraphics[width=\linewidth]{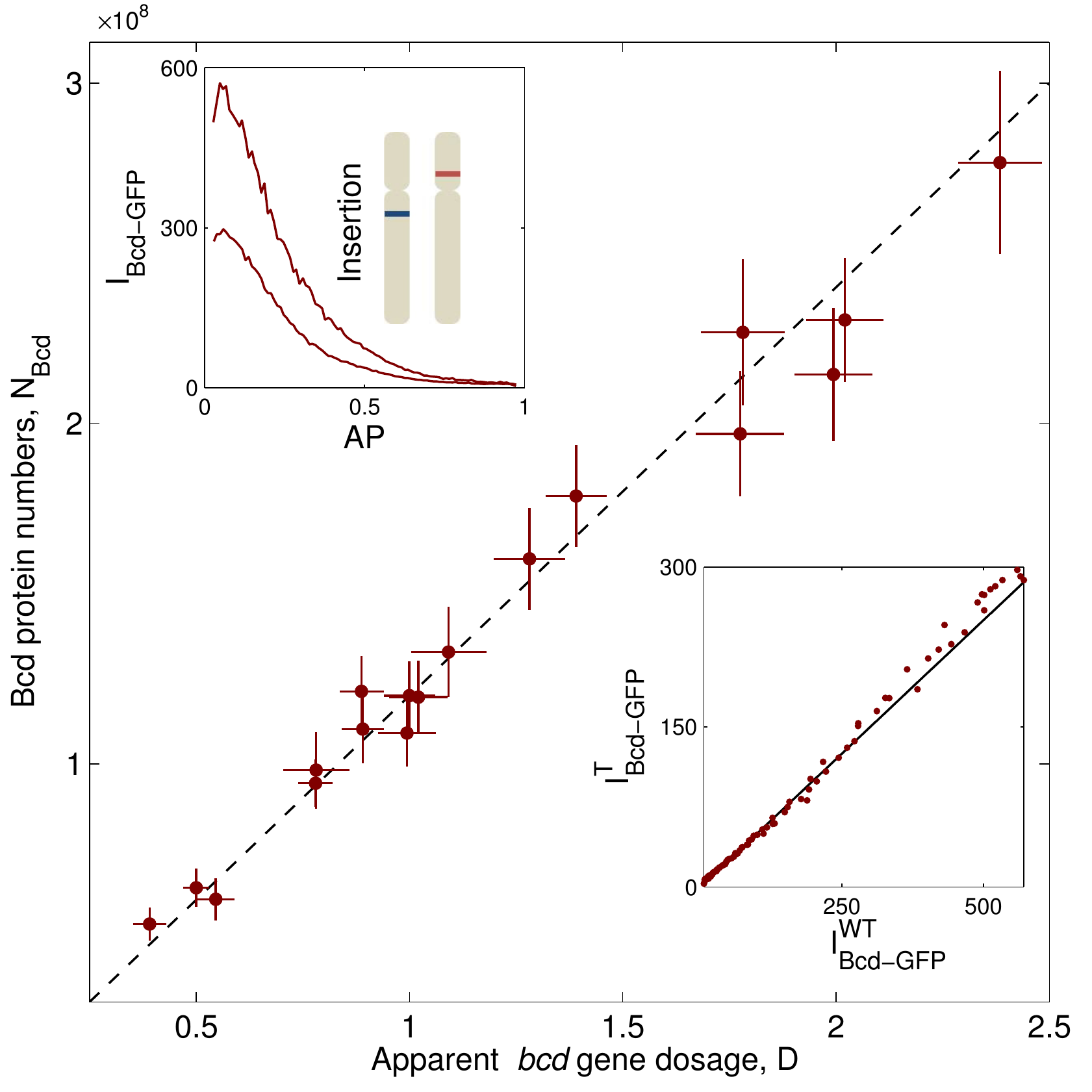}
\caption{Bcd protein counts in embryos scale linearly with the maternal {\em bcd} gene dosage. Bcd-GFP transgenes are inserted at different genome locations to generate fly lines with varying {\em bcd}  gene expression. Insertions in two such fly lines (red and blue in top inset) express at different rates due to chromosomal environmental differences which leads to quantitatively different Bcd protein intensity profiles $(I_{Bcd-GFP})$. The slope of a scatter plot of the Bcd-GFP intensity of a given (target) fly line against that from a reference line expressing Bcd-GFP at wild-type level (black line in bottom inset) measures the difference in protein concentration and is defined as the apparent {\em bcd} gene dosage {\em D}. We measure the total number of Bcd proteins in 18 fly lines with previously quantified Bcd-GFP dosage {\em D} (wild-type has {\em D}=1)~\cite{Liu:2013}. To obtain the total protein number in embryos from a given (target) fly line, their Bcd-GFP dosage {\em D} is multiplied by the ratio of their average volume to that of wild-type embryos (SI  Materials and Methods, Fig~S3). These counts are relative to the number of proteins in the reference fly line which we measure in fixed embryos using independent optical approach (see text). The total absolute protein number of Bcd-GFP molecules is obtained by multiplying these counts by the number of Bcd-GFP molecules in the reference fly line. The main panel shows the total absolute protein number $N_{\rm Bcd}$ in embryos from all 18 fly lines as the maternal  {\it bcd} gene dosage {\rm D} is varied systematically. X-axis: error bars are standard deviations in dosage measurements~\cite{Liu:2013}, Y-axis: error bars are standard deviations in protein counts relative to the reference fly line and do not include the independent error for calibrating to absolute counts in the reference fly line (SI  Materials and Methods).}
\label{fig1}
\end{figure}

Using this reference fly line we calibrate the total number of Bcd-GFP proteins in embryos from the rest of the transgenic fly line set. Specifically, embryo-to-embryo Bcd-GFP fluorescences are directly compared to measure differences in Bcd-GFP concentrations (i.e. gene dosage, Fig.~4, bottom inset) and we normalize by the ratio of their volumes to measure the relative change in total Bcd-GFP protein counts from those in the reference fly line embryos (SI Materials and Methods). The obtained total number of Bcd-GFP molecules reflects relative differences across fly lines and is stripped from the independent measurement errors in calibrating to absolute protein numbers. Using this approach, we calculate the total Bcd-GFP  counts in all 18 transgenic fly lines (Fig.~4, main panel) and observe that they scale linearly with {\em bcd} gene dosage over a range of more than two-fold changes in either direction of that of wild-type.  

Together, these observations indicate a scenario where the transition from one molecular species to another is achieved in a linear feed forward manner. Specifically, small changes in the maternal {\em bcd} gene dosage lead to proportional changes in the mRNA and protein counts. All processes  from the maternal gene expression to the protein synthesis in the embryo must operate with very high precision: $10\%$ changes in the apparent maternal gene dosage are reflected as $10\%$ changes in protein copy numbers~\cite{Gregor:2007a, Liu:2013}. Achieving such high fidelity with this linear setup is only possible if the female counts mRNA molecules precisely and individual embryos keep identical translational kinetics.

\vspace{.5cm}
\noindent {\bf mRNA Counts Are an Order of Magnitude Larger Than Expected.}  Having measured both absolute mRNA and protein counts, we take the opportunity  to relate these two quantities to the kinetic details of the Bcd translational apparatus. Previous intensity measurements of fluorescently tagged {\em bcd} mRNAs have shown that the total molecule count remains constant throughout the first two hours of development (i.e. early n.c. 14)~\cite{Little:2013}. During this period, Bcd proteins are degraded uniformly with a lifetime of $\tau\!=\!50$ min~\cite{Drocco:2011}. In the simplest model, Bcd proteins are uniformly translated and at time $t$ of expression their number is given by ${N_{\rm Bcd}(t) = k\tau M_{bcd} (1-\exp(-t/\tau))}$ where $k$ is the translation rate from mRNA to protein, and $M_{bcd}$ is the number of {\em bcd} mRNA (SI Materials and Methods). Therefore, given ${N_{\rm Bcd}=(6.7\pm1.5)\times10^7}$ molecules in approximately two hour old embryos (i.e. ${t_{14}=146\,{\rm min}}$) and ${M_{bcd}=(7.5\pm0.8)\times10^5}$ mRNA molecules, we calculate a translation rate of $k=0.03$ proteins per mRNA per second.

To gain intuition for the value of $k$, it is important to appreciate that in the early embryonic environment molecular patterns are generated extremely rapidly: the first 9 cycles of nuclear divisions last just 8 minutes each~\cite{Foe:1983}. Therefore we expect the translational apparatus to operate at near optimal rates. In particular, given a maximal translation rate of $\sim\!\!10$ amino acids per second for eukaryotic organisms under optimal growth conditions~\cite{Bonven:1979}, the $\sim\!500$ amino acid-long Bcd protein should be translated at $k_{\rm est}=0.26$ proteins per mRNA per second (with $13$ ribosomes simultaneously working on a {\em bcd} mRNA molecule~\cite{Qin:2007}). This is an order of magnitude higher than the translation rate we obtained above.

The calculation of the {\it bcd} mRNA translation rate $k$ assumes that the mRNA molecules are uniformly translated. However, there is evidence that temporaly regulated poly-adenylation of {\em bcd} mRNA can lead to a $15\text{--}30$ min delay in the activation of translation and to a time-dependent translation rate with a monotonic increase over the following 1.5h time window~\cite{Salles:1994}. 
However, even taking these effects into account, our calculation still yields a translation rate that is well below the maximally available one. Therefore, our estimations suggest that the translational apparatus performs at suboptimal translation rates and compensates by utilizing a source with a large number of mRNA molecules.

%
%
%
%

\section{Discussion}
We use an innovative combination of experimental techniques  to directly investigate the origin of reproducibility of one of the first patterns that trigger the developmental cascade in the {\it Drosophila} embryo, the Bcd morphogen gradient. We measure absolute protein and mRNA molecules in individual  embryos to trace the reproducibility of the Bcd gradient to the precision with which the female fly carries out a genetic program. Specifically, we demonstrate that during oogenesis the female assembles an mRNA source with a predetermined number of {\em bcd} mRNA molecules set at the level of {\em bcd} DNA, i.e. flies with half the number of gene copies deposit half as many mRNA. The female controls the number of deposited {\em bcd} mRNA molecules with $8\pm2\%$ precision from embryo to embryo. Identical processing of each mRNA source in each embryo leads to the observed reproducibility of the Bcd protein gradient. Thus, our results demonstrate that reproducibility is set during oogenesis. 

Transmitting molecular signals in a linear feed-forward process has a propensity for escalating noise, yet changes as low as 10\% in the apparent maternal gene dosage result in  10\% changes in the number of Bcd proteins. This observation demonstrates that the molecular network which sets up the Bcd gradient must maintain such precision at each transition from one molecular species to the next. In particular, the reproducibility of {\em bcd} mRNA counts and their proportionality to maternal copy numbers indicate that the synthesis and transport processes that lead-up to mRNA localization during oogenesis are in fact precisely controlled in different females. Analogously, the linear response in Bcd counts to changes in the maternal gene copies indicates identical {\em bcd} translation kinetics across embryos. Our results suggest that the different sources of fluctuations in the embryo must be kept at similarly low levels, in which case the Bcd profile is theoretically predicted to favor an exponential shape~\cite{Saunders:2009}. More generally, the establishment of the Bcd gradient, from the earliest stages during oogenesis all the way to when it exerts its function in the embryo, emerges as a paradigm to study how molecular networks are coordinated to maintain precision.

Our data indicates that utilizing large numbers of molecules and simple physical principles such as temporal and spatial averaging are involved in maintaining this precision.  For instance, to reproducibly assemble the mRNA source from embryo to embryo, the female counts mRNA molecules and in principle has to deal with counting (Poisson) noise. However, nearly a million {\em bcd} mRNA molecules are employed in this case, so Poisson noise is completely overridden. Moreover, this large number of mRNA molecules is generated over a period of several days during which {\em bcd} transcripts are synthesized and subsequently transferred to the embryo~\cite{Spradling:1993, Becalska:2009}. This long time span relative to the time scale of microscopic events (i.e. the synthesis of a single mRNA molecule) could naturally be utilized for time integration to generate a reproducible number of source molecules. On the other hand,  within the embryo, it takes only $\sim\!1$hr to translate the mRNA and to establish a reproducible Bcd gradient that spans the whole length of the embryo. We suggest that to achieve this reproducibly, embryos employ a lower than optimal translation rate with a large number of source mRNA molecules, facilitating spatial and temporal averaging~\cite{Gregor:2007b,Erdmann:2009}. How these interactions are matched to produce reproducible outcomes remains an open question for further investigation. 

The details of the setup of the Bcd gradient are also interesting from a developmental point of view. Functionally, the Bcd protein is a key transcription factor that controls the expression of a multitude of target genes along the entire AP axis of the embryo~\cite{Chen:2012}. Notably, Bcd concentration decreases over more than an order of magnitude from one end to the other while eliciting a response in its target genes~\cite{Gregor:2007b}. Therefore, any local inhomogeneity of Bcd protein concentration has the potential to confuse the developmental process. By depositing a large number of mRNA the female mitigates this problem:  the mRNA source populates the whole anterior pole of the embryo allowing  diffusion to smooth out the point-source-like nature of individual mRNA molecules, thus avoiding ``local" gradients. Additionally, it has been suggested that adjusting the strength of the mRNA source aids the embryo in solving scaling issues of the patterns along the AP axis~\cite{Cheung:2011}; the precise maternal control over the mRNA source coupled with the sensitivity of the Bcd gradient to small fluctuations in the number of source mRNA permit for such a scenario. Intriguingly, the high sensitivity of the Bcd gradient to maternal inputs is in contrast with the feedback-regulated robustness of the BMP gradient along the DV axis of the embryo~\cite{Eldar:2002}. This gives an opportunity to ask why each of these strategies (sensitivity vs robustness) is used in any particular situation.

Experimental work geared toward understanding how developmental networks coordinate to generate spatially reproducible morphological features has quantified the precision with which some of the molecular patterns along the developmental cascade  specify position. The Bcd gradient specifies position with a precision of one nuclear distance~\cite{Gregor:2007b} while the next patterning layer (the gap genes) collectively specifies position with precision of half a nuclear distance~\cite{Surkova:2008,Dubuis:2013}.  These observations are considered to be evidence that spatial precision increases from the initial patterns to their readouts and that this is achieved through collective feedback cross-regulation among the gap genes~\cite{Manu:2009a}. In this case, the spatial reproducibility of the final morphological structures is acquired within the embryo and one is lead to ask how do molecular networks interact to increase precision. However, this view is potentially problematic since it implies that information about position is somehow created by the embryo.

On the other hand, the macroscopic outcomes of the developmental cascade, i.e. rows of cells with identical fates, are generated with single-cell precision which is already observed in the initial setup of the Bcd gradient. Additionally, it has been shown that signals from multiple maternal inputs are integrated to determine the location of the cephalic furrow~\cite{Liu:2013}.  Here we find that the female precisely controls the assembly of the {\em bcd} mRNA source and, because all of the initial patterning signals are set up also as localized mRNA sources, it is reasonable to suggest that these cues are controlled with similar precision. The initial conditions for the first protein patterns are then specified identically from embryo to embryo and the ensuing developmental cascade acts collectively to integrate the multiple maternal signals and determine the locations of morphological structures. Hence, morphological features emerge reproducibly from one embryo to the next as a result of the precision with which the female controls the initial patterning signals. The early segmentation cascade in the {\em Drosophila} embryo is therefore a striking example of a molecular network which integrates precise initial conditions to achieve reproducible macroscopic outcomes. 

%
%
%
%
%

\section{Materials}

{\small \noindent{\bf Fly strains}: females lacking Staufen activity were generated from intercrosses of stocks carrying stauD3~\cite{Lehmann:1991} to w[1118]; Df(2R)BSC483/CyO (Bloomington Drosophila Stock Center 24987). Bicoid deficient embryos were obtained from w[1118];Df(3R)BSC422/TM6C, Sb[1] cu(1) stocks (Bloomington Drosophila Stock Center 24926).

\noindent{\bf Fluorescence in situ hybridization (FISH) and imaging}: Embryos of ages  10\--60min were fixed using $5\%$ parafomaldehyde followed by immediate \emph{in situ} hybridization as described in~\cite{Little:2011}, using n=90 Atto 565-conjugated probes complementary to the {\em bcd} open reading frame. Embryos were mounted in Vectashield (Vector Labs).  We imaged embryos in nuclear cycles 2 and 3 using a 63$\times$ HCX PL APO CS 1.4 NA oil immersion objective on a Leica SP5 laser-scanning confocal microscope with 561 nm excitation wavelength, a linear pixel dimension of 75.7 nm and z-spacing of 420 nm. Each embryo was imaged in three stacks: anterior, middle, and posterior with arbitrarily large overlapping regions; stacks were acquired from the top surface (closest to the objective) inward towards the embryo's midsaggital plane to scan half of the embryo's volume. Image analysis was performed as described in~\cite{Little:2011, Little:2013} with enhancement toward measuring {\em bcd} mRNA counts in the densely populated anterior volume (SI Materials and Methods). 

\noindent{\bf qRT-PCR and calibration curves}: Embryos of ages  10\--30 min are dechorionated in 100\% bleach for 2 min. After 1$\times$PBS rinsing,  n=1,2,4 or 8 embryos are loaded under a microscope on a custom made stainless steel tip and homogenized in thin-walled PCR tubes containing TRIzol reagent (Invitrogen), Glycogen (Ambion) and Phase Lock Gel Heavy (5 Prime).  RNA is extracted by overnight IPA precipitation and reverse transcription is carried out using SSIII (Invitrogen) using the reverse primer from the gene-specific primer pair designed for qPCR amplification (see below). Calibration curve constructions and controls are described in (SI Materials and Methods). 
 Real-time qPCR is performed on a 384-well plate using SYBR Green (Applied Biosystems) with primer pair 5' GCAAGAAGACGACGCTACAGA 3' (forward primer) and 5' AAGGCTCTTATTCCGGTGCT 3'  (reverse primer) and carried out on Applied Biosystems 7900HT Fast Real Time PCR system using  standard temperature protocol and automatic threshold detection. 
 %

\noindent{\bf Absolute qRT-PCR quantification of transcripts in embryos}:  We construct an RNA calibration curve and use it as a reference to absolute mRNA numbers in embryos under the following assumptions (i) the efficiency of RNA isolation is identical for both embryonic and synthetic RNA, (ii) extraction and reverse transcription are linear processes with total efficiency $\eta$. Therefore after undergoing RNA isolation and reverse transcription, an aliquot with $M_{\rm ref}$ reference mRNA molecules and sample of embryos with $M_{bcd}$ molecules yield samples with $\eta M_{\rm ref}$ and $\eta M_{bcd}$ cDNA molecules respectively. During qPCR, a sample with $N_{\rm cDNA}$ DNA molecules is amplified with efficiency $\varepsilon$, such that the number of DNA molecules at amplification cycle is $N_{C}=N_{\rm cDNA}\times\varepsilon^C$. Double stranded DNA is fluorescently labeled with reporter (SYBR Green) dye and the total fluorescence of the sample is linearly proportional to the number of DNA molecules $N_C$.  Setting a fluorescence threshold corresponds to having a threshold at $N_{\rm th}$ DNA molecules and allows the samples to be compared based on the number of cycles required to reach this threshold. Explicitly, by recording the cycle $C_{\rm th}$ at which the sample crosses that threshold we obtain an equation for the RNA calibration curve:
 $C_{\rm th}^{\rm ref} = -\frac{1}{\log(\varepsilon)}\log(\eta n M_{\rm ref}) + \frac{1}{\log(\varepsilon)}\log(N_{\rm th})$. An analogous equation for a sample with $n$ embryos each with $M_{bcd}$ {\em bcd} mRNA per embryo is given by: $C_{\rm th}^{\rm emb} = -\frac{1}{\log(\varepsilon)}\log(\eta n M_{bcd}) + \frac{1}{\log(\varepsilon)}\log(N_{\rm th})$. For both curves the relationship between input number of RNA molecules and $C_{\rm th}$ is linear on a log-linear plot. The two curves have identical slopes, i.e.  $S = -{1}/{\log(\varepsilon)}$, which are measured with high precision by the standard DNA curve and are separated by an offset $\Delta$ (see main text, Fig.~4A). The mean number of mRNA per embryo is found by subtracting the two equations: $M_{bcd} = M_{\rm ref} \varepsilon^{-\Delta}$. \indent The uncertainty, $\sigma_{bcd}$, in the absolute copy number is calculated using standard error propagation methods:
$\frac{\sigma_{bcd}}{M_{bcd}} = \sqrt{\frac{\sigma_r^2}{M_{\rm ref}^2} +\sigma_\Delta^2\log(\varepsilon)^2 + \frac{\sigma_\varepsilon^2}{\epsilon^2}\Delta^2}$.The terms under the root are the contributions from the uncertainty in measuring i) the number of reference RNA, ii) $\sigma_r$, the offset between the two curves, $\sigma_{\Delta}$, and iii) the slope, $\sigma_\varepsilon$. The uncertainty in measuring the copy-number of reference RNA was estimated by calibrating the absolute numbers using Bio-Rad Spectrometer and NanoDrop and obtaining the SD between the measurements; we find that $\sigma_r/M_{\rm ref}=$5\%. The error in the slope $S$ and the offset $\Delta$ are measured from the fits to the data in each individual experiment. 

\noindent{\bf Measuring absolute Bcd-GFP numbers in embryos}: Embryos expressing Bcd-GFP were fixed at $\sim$16 minutes into nuclear cycle 14 with 10\% formaldehyde and their vitalline membrane was removed by hand-peeling as described~\cite{Little:2011}. Fixed embryos were mounted in aquapolymount and imaged at their midsaggital plane with a home-built two-photon microscope using an oil-immersion 25X (NA 0.8) Zeiss objective ( PSF$_{\rm xy}$=0.5$\mu$m, PSF$_{\rm z}$=2$\mu$m). The excitation wavelength was 970 nm and the average image power at the speciment was 13mW to prevent any observable photobleaching. In fixed embryos, all fluorescent eGFP fluorophores are matured and therefore all Bcd-GFP molecules produced by the embryo are optically detectable~\cite{Little:2011}. The average Bcd-GFP concentration is measured by calibrating the mean intensity of the Bcd-GFP fluorescence to that of a 54 nM eGFP solution, a gift of H.S. Rye (Texas A\&M University).  The total number of Bcd-GFP molecules in the embryo is calculated by reconstructing, in software, the whole embryo and multiplying its volume by the average Bcd-GFP concentration (SI Materials and Methods).

\begin{acknowledgments}
We thank E. Gavis, S.C. Little, K. Sinsimer, S. Rutherford, and M. Tikhonov for advice and discussion; B. Bassler for providing the qPCR machine; and S. Blythe, E. Cox, K. Mody,  and A. Sgro for comments.  This work was supported by NIH Grants P50 GM071508 and R01 GM097275, and by Searle Scholar Award 10-SSP-274 to T. Gregor. 
\end{acknowledgments}}

\end{document}